\documentstyle[11pt,newpasp,twoside,epsf]{article}
\def\edcomment#1{\iffalse\marginpar{\raggedright\sl#1\/}\else\relax\fi}
\reversemarginpar

\begin{document}
\title{The Merging History of the Milky Way Disk}
 \author{Rosemary F.G.~Wyse}
\affil{The Johns Hopkins University, Dept.~of Physics \& Astronomy, Baltimore, MD 21218, USA}

\begin{abstract}
The stellar populations of the stellar halo and of the thick disk of
the Milky Way reveal much about the merging history that our Galaxy, a
typical large disk galaxy, has experienced.  Our current understanding, described here, 
implies a rather quiet evolution for the disk, back to redshifts of
order 2.
\end{abstract}

\section{Introduction}
Following the Copernican principle, the Milky Way is a typical galaxy,
albeit one for which we can obtain atypically detailed observational
data.  Thus galaxies like the Milky Way should form naturally in
viable models of galaxy formation and evolution.  I here discuss what
`like the Milky Way' means, in terms of the derived merging history of
the Milky Way disk.  These constraints relate to the mass assembly and
accretion history, and complement those constraints from the star
formation history of the various stellar components of the Milky Way
Galaxy (cf. Gilmore, this volume).  Interactions and
mergers clearly happen, for example the Milky Way is currently having
a close encounter with the Sagittarius dwarf spheroidal galaxy 
(Ibata, Irwin \& Gilmore
1994; 1995). 

All models of galaxy formation involve star formation in
self-gravitating regions that are smaller than the initial
`proto-galaxy', whether they reflect the small-scale power in the
primordial power spectrum of density fluctuations (e.g. in
Cold-Dark-Matter-dominated cosmologies, Blumenthal et al.~1984) or
rather are Jean-mass fragments (e.g.~Fall \& Rees 1985).  Subsequent
disruption of these systems, by internal effects such as stellar
feedback and/or by external effects such as tides, forms the field
stellar populations of galaxies we observe today.  Thus merging and
assimilation of small-scale structure is present at some level in all
models of galaxy formation, and the crucial questions are what merged,
and when did they do so?

Accretion, the acquisition and assimilation of (smaller) objects that
had an independent existence for many dynamical times, can occur in a
variety of scenarios. For example dynamical friction can cause the
orbits of massive satellite galaxies to decay (on a timescale of $\sim
(M_{galaxy}/M_{satellite}) \, t_{crossing}$).  Alternatively, the
accretion could reflect the hierarchical growth of structure set by
the initial primordial power spectrum and the cosmological
parameters. Indeed, one should remember that even `monolithic
collapse' models predict long time (continual even) infall, again 
depending on the cosmological parameters (Gunn
\& Gott 1972).  But again, the models may be distinguished by 
what is accreted, and when.

The Galactic disk is particularly relevant since disks are cold, and
susceptible to heating by merging/accretion -- thus their fragility
constrains the past merging history. If a satellite of mass
$M_{satellite}$, on an initial orbit characterised by velocity $v_s$,
can impart all of its orbital energy into a stellar disk, mass
$M_{disk}$ and internal velocity dispersion $\sigma_{disk}$, then the
disk is heated by (e.g. Ostriker 1990) $$\Delta \sigma^2_{disk} \sim
{M_{satellite} \over M_{disk}} \, v_{s}^2.$$ The actual heating
efficiency depends on many parameters, but unless the merging systems
are essentially completely gaseous, and so can radiate away a large
part of the energy, the concept of imprinting a signature of merging
is valid.  The merging history of the disk is written in its vertical
structure.

In principle one can rule out cosmological models, assuming the
`cosmic variance' is small and the Milky Way is typical, but even so,
predictions from the models for the merging rates of galaxies, as
opposed to dark haloes, remain uncertain.
Toth \& Ostriker (1992) evaluated the constraints from the disk
structure back to the formation of the Sun, some 5Gyr ago.  Here we
demonstrate how improved data can extend this back essentially to the onset of disk formation.

\section{Low Density (parts of) Satellites: the Stellar Halo}

The relative density and mass of a satellite galaxy are the important
parameters that determine its effect on the Galactic disk.  Low density, `fluffy', satellites
will be tidally disrupted at large Galactocentric radius, and,
assuming a typical orbit (cf.~Moore et al.~1999a), their stars 
will contribute to the stellar
halo and their gas to the disk, modulo angular momentum.  Further, the
outer, low-density, regions of satellites will contribute to the halo,
while their higher density core regions could be accreted into the
disk.  Thus evidence for accretion into the halo is arguably
necessary, but of course not sufficient, for accretion into the disk.

What evidence is there for accretion into the stellar halo?  As
described further by Gilmore (this volume), the age distributions both
the field stars and the globular clusters of the halo show a dominant
old population, with at most 10\%, biased to the more metal-rich
tracers, being candidates for `intermediate-age'. 
Thus if the accretion of
satellite galaxies is the dominant process of halo formation
(e.g.~Helmi \& White 1999; Bullock, Kravtsov \& Weinberg 2000), the
satellites must be accreted when they contain only old, metal-poor
stars (although note that even 5\% of the stellar halo is several tens
of the present dwarf spheroidal companion galaxies).  As discussed
further by Gilmore, the star formation histories of the
existing/surviving low surface brightness dwarf companions to the
Milky Way are varied, but rarely is there the case of a very
short-lived, early burst of star formation.  Thus assuming that a
dwarf galaxy left to itself would form stars over an extended period,
as did the typical surviving dwarf, then any accretion is restricted
to have occurred only very early (Unavane et al.~1996).  Plausibly the
putative satellites were quite gas-rich at those early epochs, and
would indeed contribute to growth of the gaseous disk.

Phase space signatures of accretion would persist longest in the outer
halo, where all timescales are longer (e.g. Johnston, Hernquist \&
Bolte 1996). It should be noted that the bulk of the mass of the
stellar halo is located interior to the solar circle; for typical
models of the density profile of the stellar halo (e.g.~Carney, Latham
\& Laird 1990; Morrison 1993), the halo exterior to 8~kpc is $\sim 2
\times 10^8L_\odot$ and is around 20\% of stellar halo interior to
8~kpc. The mass exterior to the periGalacticon of the Sagittarius
dwarf spheroidal galaxy, $r \sim 16$~kpc (Ibata et al.~1997),
corresponds to $\la 10^8L_\odot$, somewhat less than ten times the
present luminosity of the dSph itself (providing a real constraint on
models for the disruption of the Sgr dSph, which can contravene this
limit).

Early results from the imaging part of the Sloan Digital Sky Survey
have shown the immense wealth of information in a uniform, wide-area
survey.  Structures in the outer halo have been identified from
non-uniform A-star counts, with as much as $6 \times 10^6M_\odot$ in a
structure out at $r \sim 45$~kpc (Yanny et al.~2000), and in counts of 
RR~Lyrae
stars (Ivezic et al.~2000).  These may be streams from the Sgr dSph
(Ibata, Irwin, Lewis \& Stolte 2000), but one should note that
`globular clusters' could have contributed a significant stellar mass
to the halo (cf. Fall \& Rees 1977; Gnedin \& Ostriker 1997).  Indeed,
this is an on-going process, with `tidal tails' detected around at
least twenty of the present-day globular clusters (Leon, Meylan \&
Combes 2000).  Thus one expects many tidal streams of old, metal-poor
stars in the halo, which have nothing to do with `accretion' or
cosmological structure formation.

Accreted stars (and disrupted globulars) may also be observable through kinematics, as moving
groups.  And indeed a metal-poor, old (typical of the stellar halo)
`debris stream' has been identified in the solar neighborhood (Helmi
et al.~1999) that may contain, by mass, several percent of the stellar
halo outside of the solar circle, or $\la 10^7M_\odot$ (based on $\sim 10$ stars). The broad
metallicity distribution (cf. Chiba \& Beers 2000) argues against an
origin in a `proto-globular cluster', and supports a dwarf galaxy.  More data are clearly warranted.    

The surviving remnants/cores/systems of tidal disruption will act on, and 
contribute to, the disk if they are massive enough.

\section{Massive, Dense Satellites: the Thick Disk}

\subsection{Simulations}

There have been many simulations of the effect on a pre-existing thin
stellar disk of a `minor merger'.  As displayed clearly in Figure~11
of Quinn \& Goodman (1986), their `standard model', a massive and
dense satellite on a prograde circular orbit inclined moderately
(45$\deg$) to the plane of the disk, couples efficiently into the
vertical motions of the disk, and the $z-$motions of the satellite are
rapidly damped, followed by subsequent spiralling in of the satellite,
in the plane of the disk.  The satellite loses material as it spirals
in and its effective tidal radius decreases, and it heats the disk as
it gives up orbital energy.  The heating is a combination of local
effects, plus resonant excitation of large-scale bending waves
(Sellwood, Nelson \& Tremaine 1998).  The heating amplitude depends
on many parameters of the initial orbit of the satellite, such as 
the inclination, the pericenter, amplitude and sense of angular
momentum, on its
internal structure -- density profile and  mass -- 
 and on the coupling between the
various components of the interacting systems (e.g. Walker, Mihos \&
Hernquist 1996; Huang \& Carlberg 1997; Velazquez \& White 1999).  
A robust satellite of around 20\% of the present mass of the stellar 
disk could  create a thick disk with a scale-height of around 1kpc. 

However, as we will see below, the extant simulations are not
appropriate to model the formation of the Galactic thick disk.
Rather, we need simulations that better match the early stages of disk
evolution. At a minimum, gas -- which of course can cool after being
heated -- must be included, both in the satellite and in the disk.

\subsection{The Real World}

\begin{figure}[!h]
\plotfiddle{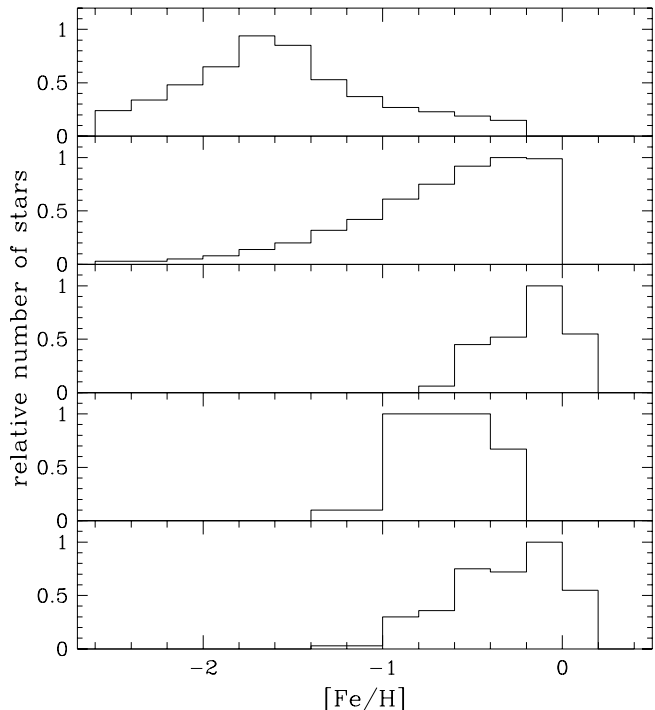}{3.5in}{0}{45}{45}{-150}{-30}
\hskip -0.75truecm
\caption{The metallicity distributions of 
representatives of the stellar populations of the Milky Way Galaxy.
Where possible, a measure of the true iron abundance is plotted.  The
panels are (top to bottom) the local stellar halo (Carney et al.\/ 1994,
their kinematically-selected sample); the outer bulge K-giants (Ibata
\& Gilmore 1995), truncated at solar metallicity due to calibration
limitations; the volume-complete local thin disk F/G stars (derived
from the combination of the Gliese catalogue and {\it in situ\/}
survey); the volume-complete local thick disk F/G stars (derived
similarly); and lastly the `solar cylinder', i.e. F/G stars integrated
vertically from the disk plane to infinity. This figure is based on
Fig.~16 of Wyse \& Gilmore (1995).}
\end{figure}

The Galactic thick disk was first detected through star counts at high
latitude (Gilmore \& Reid 1983), although surface photometry of
external S0 galaxies had earlier revealed `thick disks' in them
(Burstein 1979; Tsikoudi 1979).  Many subsequent star count analyses
have shown that a thick disk component is required in addition to the
standard thin disk and stellar halo (e.g.~Buser, Rong \& Karaali 1999;
Phleps et al.~2000).  The structural parameters for the thick disk are
rather uncertain using star count data alone, and it is extremely
important to determine the global structure and mass of the thick
disk.  Available constraints from star counts suggest a scaleheight
about a factor of four greater than that of the thin disk, a local
normalisation of around 2\% (these two are anti-correlated in the
analyses) and a scale-length equal to that of the thin disk.  The mass
of the thick disk is then around 10\% of the mass of the thin disk,
although some determinations as high as 20\% are also consistent.

The astrophysical parameters of the thick disk reveal its evolutionary status. 
At the time of its discovery, it was postulated that the thick disk
was formed by local compression of the stellar halo by the potential
of the thin disk (Gilmore \& Reid 1983).  This was soon disproven
(Gilmore \& Wyse 1985) by the determination of distinct metallicity
distributions of thick disk and stellar halo (see Fig.~1 here).
The thick disk is also distinct from the thin disk in terms of
kinematics, seen most clearly using the vertical velocity, $W$.  Fig.~2 
here plots the 
 rank number of a star 
in iron abundance versus the sum of the absolute value of the vertical
velocity of the stars up to and including that rank, for the sample of
local F stars of Edvardsson et al.~1993
(iron abundances are obtained from echelle spectra, and the kinematics
are from the combination of radial velocities and proper motions).
For a Gaussian velocity distribution, the
slope of this plot is the value of the velocity dispersion.  
The distinct metallicity distribution of Fig.~1 leads to the
expectation that the thick disk should be dominant at iron abundances
less than about $-0.4$~dex. And indeed, the
rank corresponding to this metallicity is just where one detects a
break in the kinematics, specifically an increased velocity
dispersion.  Essentially all kinematic studies of the thick
disk find that its vertical velocity dispersion is $\sim 40$~km/s
(e.g.~review of Majewski 1993), which fits with a scale-height of
around 1~kpc, in the vertical potential derived from other samples
(e.g. Kuijken \& Gilmore 1989).

\begin{figure}[!ht]
\plotfiddle{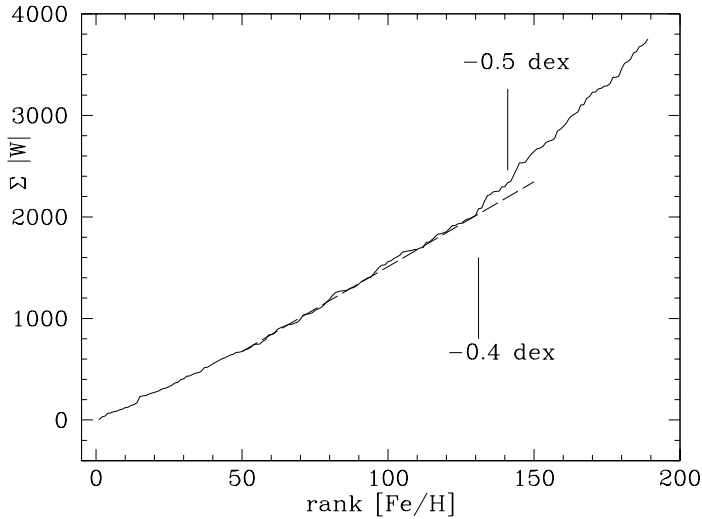}{3in}{270}{35}{35}{-150}{225}
\vskip -0.7truecm
\caption{The rank number of a star 
in iron abundance versus the sum of the absolute value of the vertical
velocity of the stars up to and including that rank, for the sample of
Edvardsson et al.~(1993) (taken from Wyse \& Gilmore 1995).  
For a Gaussian velocity distribution, the
slope of this plot is the value of the velocity dispersion.  There is
a clear change of slope at the metallicity at which the thick disk
dominates over the thin disk, indicating distinct kinematics.}
\end{figure}

 The
vertical velocity dispersion of the thick disk, $\sim 40$~km/s, is too
high to result from the `normal' thin-disk heating processes of
scattering by disk density perturbations such as giant molecular
clouds or transient spiral structure, as these saturate at lower
values of the velocity dispersion (e.g.~Spitzer \& Schwarzschild 1951;
Lacey 1991; Quillen \& Garnett 2000).  An alternative to a `heating'
mechanism is a cooling mechanism, with the possibility that the thick
disk formed during the dissipative settling of gas into the proto-disk
(Wyse \& Gilmore 1988; Burkert, Truran \& Hensler 1992).  The distinct
nature of the thick disk in these models results from the sensitivity
of star formation rate and cooling to a parameter, for example
metallicity (Wyse \& Gilmore 1988).  Such a cooling scenario would,
however, predict a fairly uniform population of thick disks, which is
not observed (e.g. Morrison 1999).

The analogous plot to Fig.~2 that employs rank in {\it age\/} within
this sample rather than in iron abundance shows similar behaviour
(Freeman 1991), in that there is a clear change in kinematics at some
rank, and that rank corresponds to an age of $\ga 12$~Gyr (the ages
are derived from Stromgren photometry, and are based on comparisons
with the VandenBerg (1985) isochrones, but Ng \& Bertelli (1998) find
little changes with use of later isochrones, or with {\sl Hipparcos}
distances).  This limit for the youngest stars in the thick disk is
also seen in Fig.~3, where again one finds few stars in the thick disk
that are younger than the globular cluster 47~Tuc (given the
calibration uncertainties in absolute ages, the relative age compared
to a globular cluster is the most robust means of stating the age
limits on the thick disk).

This old age for the thick disk, at least probed within several kpc of
the solar circle, is in agreement with the analyses of several other
samples and groups (e.g.~Wyse \& Gilmore 1985; 1988; Carney, Latham \&
Laird 1989; Fuhrmann 1998).  The spread in ages older than this limit
is poorly constrained by colours, but the age distribution of the
Edvardsson et al.~sample continues to extremely old ages.  The pattern
of elemental abundances in the thick disk can also constrain the
duration of star formation, but there is at present no concensus (compare 
Prochaska et
al.~2000 and  Gratton et al.~2000).  

The lack of young stars in the thick disk is a very significant
constraint on its formation, as we discuss below.  A robust 
age distribution is a definite {\it desideratum}.

\begin{figure}[!ht]
\plotfiddle{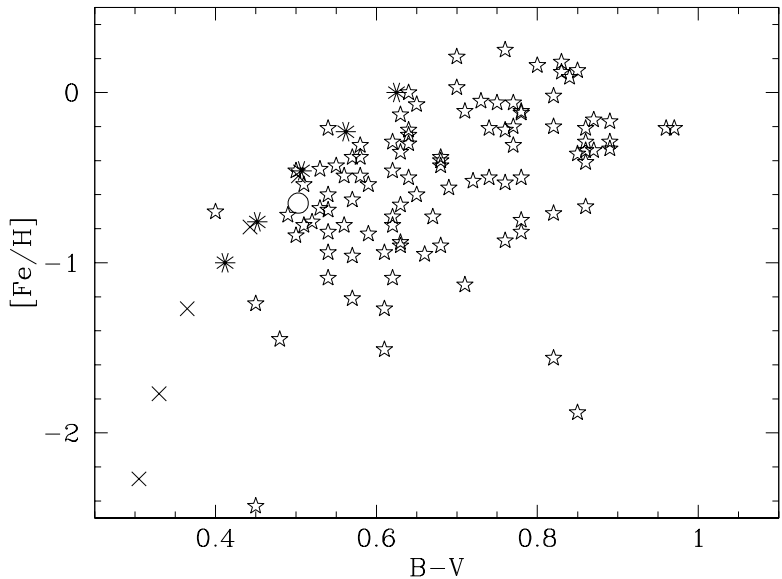}{3.25in}{270}{40}{40}{-155}{240}
\caption{Scatter plot of iron abundance {\it vs\/} B$-$V colour for
thick disk F/G stars, selected {\it in situ\/} in the South Galactic Pole 
at 1-2kpc above the
Galactic Plane (stars), together with the 14~Gyr turnoff colours
(crosses) from VandenBerg \& Bell (1985; Y=0.2) and 15~Gyr turnoff colours
(asterisks) from VandenBerg (1985; Y=0.25).  The open circle represents the
turnoff colour (de-reddened) and metallicity of 47~Tuc (Hesser et
al.~1987).  The vast majority of thick disk stars lie to the red of
these turnoff points, indicating that few, if any, stars in this
population are younger than this globular cluster.  This figure is
based on Fig.~6 of Gilmore, Wyse \& Jones (1995).}

\end{figure}

Thus the Galactic thick disk is plausibly the result of heating of a
pre-existing thin disk during the process of a minor merger.  The
`smoking-gun' evidence in support of this would be identification of
the stars now part of the Galaxy, that were initially in the satellite
responsible for the heating.  We (Gilmore, Wyse, Norris \& Freeman)
have an on-going survey of thick disk/halo stars using the `2 Degree
Field' spectrograph on the Anglo-Australian Telescope to analyse 
the overlapping halo/thick disk populations and constrain the phase space 
structure of the stripped satellite. 

\section{Implications}

The properties of the thick disk discussed above, and in
particular the conclusion that the favoured formation mechanism for
the generation of the thick disk is through a minor merger, have major
implications for the formation and evolution of disk galaxies.  

First, the fact that the youngest stars in the thick disk are
essentially as old as the globular cluster 47~Tuc (some 12.5~Gyr with
the most recent isochrones and distance calibrations; Carretta et
al.~2000) limits the last significant merger event to have occurred a
long time ago.  The limit arises due to the fact that star formation
in the thin disk has been fairly continuous since its onset, albeit
with the amplitude of star formation varying radially; for example,
derived star formation histories in the local thin disk show an
overall decline of a factor of a few from the earliest times, with
superposed `bursts' of amplitude of a few (e.g.~Rocha-Pinto et
al.~2000; Gilmore, this volume).  A merger at any given
time would heat the thin disk stars formed up to that time, and thus a
later merger would create a thick disk which contains younger stars.
Hence the Milky Way cannot have undergone a significant -- some 20\%
by mass from the extant simulations and analyses -- merger for a very
long time.  Adopting 11~Gyr as a fiducial look-back time for this
epoch (cf.~the oldest stars in the {\sl Hipparcos} sample of the thin disk,
Binney, Dehnen \& Bertelli 2000), then with the `standard'
cosmological parameter values of $\Omega_\Lambda = 1 - \Omega_{matter}
= 0.7$, $H_o = 65$km/s/Mpc, there has been no significant merger since
a redshift of $ z_{thick\, disk} \sim 2$.

Of course, there are many uncertainties in this argument, but it is an
important enough conclusion that significant effort needs to be
expended to address those areas of uncertainty.  Accepting  the merger
origin of the thick disk, these uncertainties include the existence
and value of the critical mass (and density) of a minor merger to
provide sufficient heating, the age distribution of the thick disk
(both locally and globally) and indeed the properties of the thick
disks of external galaxies. 

A second major conclusion within this scenario is that a fairly
massive stellar thin disk was already in place at the epoch of this
last significant merger, to provide the thick disk as observed now,
perhaps as much as 20\% of the present thin stellar disk, or of order
$10^{10}M_\odot$.  This early thin disk extended at least out to the
solar Galactocentric radius.  There were certainly extended disks at redshifts
significantly above unity (cf.~Somerville, this volume).  While there
is an obvious need for a robust determination of the global structure
of the thick (and thin!) disk, this conclusion is consistent with the
analysis of Brinchmann \& Ellis (2000) of field galaxies at redshifts
out to unity (see also Ellis, this volume), which concluded that the
bulk of star formation had already occurred by a redshift of 1.  This
of course rules out scenarios that propose delayed infall of
proto-disk gas, until after merging is complete, or redshifts of
unity, as the solution to the angular momentum problem in
hierarchical-clustering pictures for disk galaxy formation (e.g.~Weil,
Eke \& Efstathiou 1998).

As a corollary to this, one can envisage a bulge-disk connection that
provides an explanation for the near equality in age between the
central bulge (Ortolani et al.~1995; Feltzing \& Gilmore 2000) and the
thick disk, by positing that the bulge formed from rapid star
formation in gas driven inwards by the torquing inherent in the
merging process, perhaps accompanied by bar formation (e.g.~Noguchi
1988; Hernquist \& Mihos 1995) and globular cluster formation.  This
would also produce a correlation between bulges and thick disks, which
may indeed by the case (e.g.~Morrison 1999).

Further, the apparent unimportance of merging
in the history of the bulk of the baryonic mass of the Milky Way, combined with the angular momentum
problem for disks in general (e.g.~Navarro \& Steinmetz 1997;
Steinmetz this volume), the over-prediction of surviving satellite
galaxies (Moore et al.~1999a; Klypin et al.~1999), and the steep inner
halo density profiles (Moore et al.~1999b, but see 
Navarro, this volume) perhaps point to a
problem with the underlying Cold-Dark-Matter power spectrum.
Suppression of the small scale power can provide a solution to many of
these problems (e.g.~Sommer-Larsen \& Dolgov 1999; Moore et al.~1999a).

\medskip 
{\it Acknowledgements}: I am indebted to the conference organisers for
their financial support,  enabling me to participate in this highly
enjoyable and stimulating conference.

\end{document}